\documentclass[prb,11pt,superscriptaddress]{revtex4-1}

\usepackage{amsmath}
\usepackage{amssymb}
\usepackage{graphicx}
\usepackage{color}
\usepackage{float}
\usepackage{caption}
\usepackage{subcaption}
\usepackage{braket}
\usepackage{hyperref}
\hypersetup{colorlinks=true}
\setcitestyle{square}
\begin{document}

\title{Probing excitations in insulators via injection of spin-currents}
\author{Shubhayu Chatterjee}
\affiliation{Department of Physics, Harvard University, Cambridge MA 02138}
\author{Subir Sachdev}
\affiliation{Department of Physics, Harvard University, Cambridge MA 02138}
 \affiliation{Perimeter Institute for Theoretical Physics, Waterloo, Ontario N2L 2Y5, Canada}

\date{\today}

\begin{abstract}
We propose a spin transport experiment to measure the low-energy excitations in insulators with spin degrees of freedom, with a focus on detecting ground states that lack magnetic order. A general formalism to compute the spin-current from a metal with a non-equilibrium distribution of spins to an insulator is developed. It is applied to insulating states with and without long range magnetic order, and salient features in the spin-conductance are noted.  
\end{abstract}

\maketitle

\tableofcontents

\section{Introduction}
Observation of fractionalized excitations in insulating spin-systems has been a long-sought goal in physics. Such quantum spin liquid states, if realized in nature, would be a new quantum phase of matter with exotic properties. Certain candidate materials have strong experimental evidence for exhibiting spin liquid ground states. For example, thermal conductivity experiments on insulating frustrated triangular lattice organic salts by M. Yamashita {\em et al.\/}\cite{yamashita2009thermal} indicate presence of mobile gapless excitations. Inelastic neutron scattering experiments on single crystals of Herbertsmithite, a kagome lattice spin-half Heisenberg antiferromagnet by Han {\em et al.\/}\cite{han2012fractionalized} provide evidence for the presence of a continuum of fractionalized spinon excitations. Numerical studies on the triangular\cite{zhutriangularj1j2,Sheng15} and kagome\cite{Sheng14} lattice Heisenberg models also indicate the possibility of spin liquid ground states in certain parameter-regimes.

In spite of promising evidence for observation of spin liquids from several experiments\cite{PhysRevLett.99.137207,yamashita2008thermodynamic,yamashita2009thermal,han2012fractionalized}, the exact nature of experimentally realized ground states, and in particular, the presence of a spin-gap is still unclear. In this paper, we propose a transport experiment which can probe the mobile spin-carrying excitations of the system at low energies; these experiments are similar in spirit to those discussed recently 
by Takei {\em et al.} \cite{PhysRevLett.112.227201,PhysRevB.90.094408,Takei3} and collaborators \cite{PhysRevLett.108.246601,PhysRevB.91.140402,takahashi2010spin}
for materials with magnetic order.
Recent advances in spintronics\cite{jungwirth2012spin,spinhallSinova} have made it possible to create a spin-accumulation at boundaries of metals via the spin Hall effect. We propose to use this non-equilibrium accumulation of spins to inject a spin-current into an insulating state with spin-degrees of freedom. The spin-current is a function of the spin-accumulation voltage in the metal. Therefore, by measuring the spin-current as a function of this voltage, and looking at thresholds and exponents, we can comment on the presence of spin-gaps and the low-energy dispersion of the fractionalized spin-half excitations.

The rest of the paper is organized as follows. In section \ref{formalism}, we describe the geometry of our setup, and develop a formalism to evaluate the spin-current injected into a magnetic insulator from a metal. In section \ref{af}, we apply the formalism to evaluate the spin-current into an antiferromagnet with collinear Neel order. In section \ref{non-mag}, we first analytically calculate for the spin-current into insulating states with no long range magnetic order, including both valence bond solid states and spin liquid states. Then we go beyond the analytical approximations, and numerically identify some broad features in the spin-conductance for a spin liquid ground state\cite{PhysRevB.45.12377} on the kagome lattice, which is a candidate state for Herbertsmithite\cite{han2012fractionalized,punk2014topological}. Details of relevant calculations are contained in the appendices.  

\section{Formalism to evaluate spin-current}
\label{formalism}

\subsection{Generation and detection of spin-current}

We begin with a brief discussion of the spin Hall effects, which we shall use to generate and detect spin-currents, and then describe the exact geometry of spin injector and detector we use. A charge current  passed through a paramagnetic material can drive a transverse spin current in presence of strong intrinsic spin-orbit coupling or skew-scattering by spin-orbit coupled disorder\cite{DyakonovPerelSH,PhysRevLett.83.1834,PhysRevLett.85.393,PhysRevLett.92.126603}. The spin current impinging on the boundary is given by $J_{S} = \frac{\hbar}{2e} \theta_{SH} J_C$, where $J_C$ is the charge current density and $\theta_{SH}$ is the spin-Hall angle, and sets up a spin-accumulation at the boundary, that has been measured in experiments for both metals\cite{kimura2007room,kajiwara2010transmission} and semiconductors\cite{kato2004observation,PhysRevB.72.245330,valenzuela2006direct}. The reciprocal process, where injecting a spin current into a spin-orbit coupled paramagnetic material sets up a charge current (or voltage) transverse to the spin-current - the inverse spin Hall effect, has also been observed\cite{valenzuela2006direct,saitoh2006conversion,kimura2007room}. Furthermore, both processes have been used simultaneously to transmit electrical signals across a magnetic insulator\cite{kajiwara2010transmission}. Theoretical predictions for the spin superfluid transport through a ferromagnetic\cite{PhysRevLett.112.227201} and antiferromagnetic\cite{PhysRevB.90.094408} insulator sandwiched between two metallic reservoirs have been worked out in the linear response regime. Taking phenomenological Gilbert damping into account, the spin current density $J^{r}_{S} $ pumped into the right reservoir as a function of the spin accumulation voltage $V$ is given by\cite{PhysRevLett.112.227201,PhysRevB.90.094408} 
\begin{equation}
J_{S}^{r} = \frac{V}{4\pi}\frac{g_{l}^{\uparrow \downarrow} g_{r}^{\uparrow \downarrow}}{ g_{l}^{\uparrow \downarrow} + g_{r}^{\uparrow \downarrow} + g_{\alpha}}
\end{equation}
where $g_{l(r)}^{\uparrow \downarrow} $ is the spin flip conductance at the left (right) interface, and $g_{\alpha}$ quantifies the loss in spin current due to Gilbert damping. 

\begin{figure}[h!]
 
\begin{subfigure}{0.45\textwidth}
\includegraphics[width=0.8\linewidth, height=4.2cm]{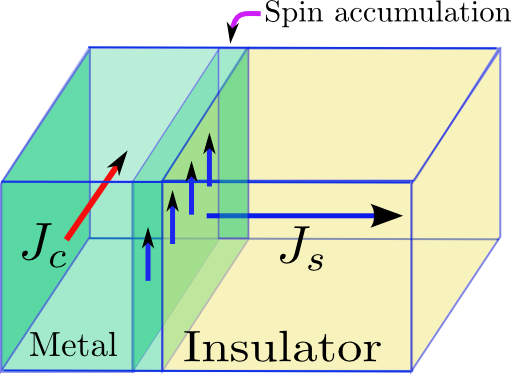} 
\caption{Spin accumulation via the spin Hall effect, and injection at the left interface}
\label{fig:subim1}
\end{subfigure}
\begin{subfigure}{0.5\textwidth}
\includegraphics[width=1.0\linewidth, height=4.2cm]{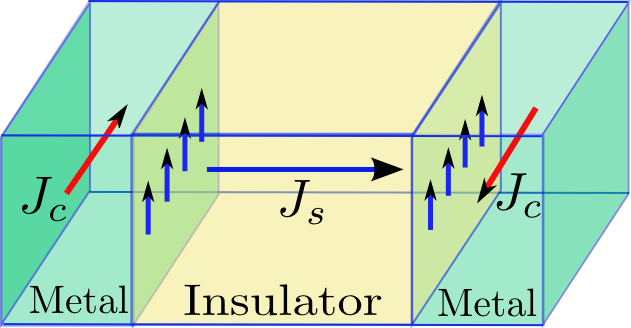}
\caption{Spin-current detection via the inverse spin Hall effect in the right metallic reservoir}
\label{fig:subim2}
\end{subfigure}
\caption{Geometry for generation and detection of spin-current}
\label{fig:genDet}
\end{figure}

Let us consider an analogous geometry, where an insulating block with spin degrees of freedom is placed in between two metallic reservoirs, as shown in Fig.~\ref{fig:genDet}. A charge current in the left metallic reservoir, in presence of strong spin-orbit coupling, will create a non-equilibrium accumulation of spin at the metal-insulator boundary. We assume that there are no thermal gradients, and that the spin accumulation can be well modeled by different chemical potentials $\mu_{\uparrow}$ and $\mu_{\downarrow}$ in the Fermi Dirac distribution at temperature $T$ for the spin-up and spin-down electrons. The left metal reservoir will subsequently relax by sending a spin-current into the spin insulator. We assume negligible loss of spin-current inside the insulator, so that the spin-current sets up a spin-accumulation at the insulator-metal boundary on the right. If the metallic reservoir on the right was initially in thermal equilibrium at $T$, the accumulated spin density at the boundary will drive a charge current via the inverse spin Hall effect. This charge-current, or the associated voltage can be detected, and therefore we can find the spin-current by measuring charge currents (or voltages) in both metallic reservoirs.

\subsection{General expression for spin-current} 
Let us choose $x$ as the longitudinal direction which is normal to the interfaces, and $z$ as the spin-quantization axis. We shall evaluate the spin-current crossing the left metal-insulator interface when $V = \mu_{\uparrow} - \mu_{\downarrow} > 0$. To make analytical progress, we assume a clean interface between the metal and the insulator, with translational invariance in the plane of the interface. The metallic reservoir is assumed to be a Fermi liquid with quadratic dispersion and Fermi energy $\epsilon_F$, so that $n_{\sigma}(\epsilon) = \left( e^{\beta(\epsilon_{\vec{k}} - \mu_{\sigma}) }+ 1\right)^{-1}$ with $\epsilon_{\vec{k}} = \frac{\vec{k}^{\,2}}{2m}$ (setting $\hbar = 1$). We shall always work in the regime where $T, V \ll \epsilon_F$, and henceforth set $\mu_{\uparrow} = \mu$, so that $\mu_{\downarrow} = \mu - V$, to simplify notations.

We assume that the electron spin $\vec{S}_e$ in the metal interacts with the boundary spins of the insulator, located at interface lattice sites $\vec{X}_j$, via a local spin-rotation symmetric local Hamiltonian
\begin{equation}
H_{int} = J \sum_{j} \vec{S}_e \cdot \vec{S}_j \; \delta(\vec{x}_e - \vec{X}_j) 
\end{equation}
Let the insulator have exact eigenstates $\{ \ket{n} \}$, then its initial state is described by the equilibrium density matrix $\sum_{n} \frac{e^{-\beta E_n}}{Z} \ket{n}\bra{n}$. For the metal, periodic boundary conditions in a large box of volume $\mathcal{V} = L_x \mathcal{A}_{\perp}$ is assumed, where $\mathcal{A}_{\perp}$ is the interface area. We now use Fermi's golden rule to calculate the rate of scattering of a right-moving electron state $\ket{\vec{k}_1, \uparrow}$ to a left-moving electron state $\ket{\vec{k}_2,\downarrow}$. The matrix element for scattering to a final state $\ket{m}$ of the insulator is given by 
\begin{equation}
\bra{\vec{k_2}, \downarrow; m} H_{int} \ket{\vec{k_1}, \uparrow; n} =  \frac{J}{2 \mathcal{V}}  \sum_{j} e^{i\vec{q} \cdot \vec{X}_j} \bra{m}  S_j^{+} \ket{n}  \: , \: \text{ defining }\vec{q} = \vec{k}_1 - \vec{k}_2
\end{equation}
Defining $\omega(\vec{k}_1, \vec{k}_2) = \epsilon_{\vec{k}_1, \uparrow} - \epsilon_{\vec{k}_2,\downarrow}$ as the energy transfer, the rate of scattering $R$ is 
\begin{eqnarray}
R &=& 2\pi \sum_{m,n}  \frac{1}{Z}e^{-\beta E_n} \left| \bra{\vec{k}_1,\uparrow; n}  H_{int} \ket{\vec{k}_2,\downarrow; m} \right|^2 \delta \left(E_n + \epsilon_{\vec{k}_1,\uparrow} - E_m - \epsilon_{\vec{k}_2,\downarrow} \right) \notag \\
& = & \frac{\pi J^2}{2 L_x^2 A_{\perp}} S_{-+}\left(\vec{q}_{\perp},\omega = \frac{2\vec{k}_1\cdot \vec{q} - \vec{q}^{\,2}}{2m}\right) 
\end{eqnarray}
where $S_{-+}(\vec{q}_{\perp},\omega) $ is the dynamic spin structure factor of the insulator at the interface, defined as
\begin{equation}
 S_{-+}(\vec{q}_{\perp},\omega) = \frac{1}{A_{\perp}}  \sum_{l,j} e^{-i\vec{q}_{\perp} \cdot ( \vec{X}_l - \vec{X}_j)}  \int_{-\infty}^{\infty} dt \; e^{i\omega t}  \langle S_l^{-}(t) S_j^{+}(0)  \rangle_{\rm thermal}
\end{equation}
The spin-current crossing the boundary for this scattering event is $\frac{q_x}{2m}$. If we have $R$ such events per unit time, then the net spin-current crossing the boundary is just $ \frac{q_x R}{2m}$. Summing over all initial electron and final states consistent with phase space constraints, the current $I_{spin,\uparrow}$ due to up-spin electrons getting reflected to down-spin ones is 
\begin{equation}
I_{spin,\uparrow} = \frac{\pi J^2 A_{\perp}}{4 m}\int_{k_{1x}>0} \frac{d^dk_1}{(2\pi)^d} \int_{q_x > k_{1x}} \frac{d^dq}{(2\pi)^d} n_{\uparrow}(\epsilon_{\vec{k}_1}) \left(1 - n_{\downarrow}(\epsilon_{\vec{k}_1- \vec{q}}) \right) \: q_x \; S_{-+}\left(\vec{q}_{\perp}, \omega = \frac{2\vec{k}\cdot\vec{q}-\vec{q}^2}{2m}\right)
\label{genIspinUp}
\end{equation}
At non-zero $T$, the reverse process where spin-down electrons get reflected to spin-up ones contribute analogously a spin-current $I_{spin,\downarrow}$ given by
\begin{equation}
I_{spin,\downarrow} = \frac{\pi J^2 A_{\perp}}{4 m}\int_{k_{1x}>0} \frac{d^dk_1}{(2\pi)^d} \int_{q_x > k_{1x}} \frac{d^dq}{(2\pi)^d} n_{\downarrow}(\epsilon_{\vec{k}_1}) \left(1 - n_{\uparrow}(\epsilon_{\vec{k}_1- \vec{q}}) \right)\: q_x \; S_{+-}\left(\vec{q}_{\perp}, \omega = \frac{2\vec{k}\cdot\vec{q}-\vec{q}^2}{2m}\right)
\end{equation}
The net spin-current is therefore given by the difference of the two contributions listed above
\begin{equation}
I_{spin} = I_{spin,\uparrow} - I_{spin,\downarrow}
\label{spinc}
\end{equation}
 
\subsection{Simplifications for certain physically relevant structure factors}
\label{simplify}

\begin{figure}[!h]
\centering
\includegraphics[scale=0.6]{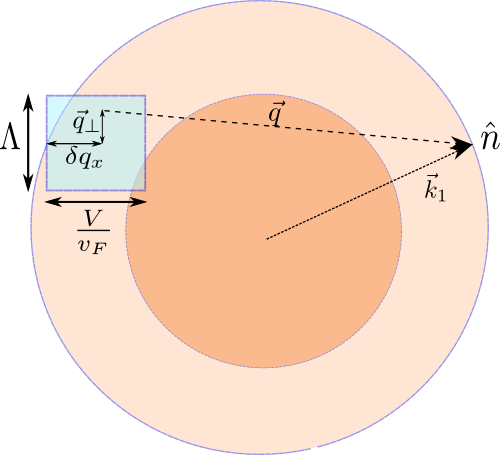}
\caption{Allowed phase space for scattering of an electron with given initial momentum}
\label{fig:phaseSpace}
\end{figure}

The expression for the spin-current can be considerably simplified once we note that at $T \rightarrow 0$, scattering is essentially restricted within an energy window of $V$. For $\omega \lesssim V$, we assume that the dynamic structure factor $S_{+-}(\vec{q}_{\perp}, \omega)$ assumes large values only for small $|\vec{q}_{\perp}|$. This is physically relevant for several systems where excitations at large momenta typically have large energy cost. As Fig.~\ref{fig:phaseSpace} shows, if the system does not have excitations at $\omega \lesssim V$ for $|\vec{q}_{\perp}| \gtrsim \Lambda$, then scattering is restricted within a patch of dimensions $\frac{V}{v_F}\times \Lambda^{d-1}$, $v_F$ being the Fermi velocity. 

To exploit this, we approximate the initial momentum $\vec{k}_1 \approx k_F \, \hat{n}$, and linearize the energy transfer $\omega$ about the point of elastic scattering as follows
\begin{eqnarray}
\vec{q} &=& 2k_F (\hat{n}\cdot\hat{x}) - \delta q_x \hat{x} - \vec{q}_{\perp} \notag \\
\omega(\vec{k}_1,\vec{q}) &=& v_F \left[  (\hat{n}\cdot\hat{x})  \delta q_x - \hat{n}\cdot \vec{q}_{\perp} \right] + \mathcal{O}(\delta q_x^2, q_{\perp}^2)
\label{linearize}
\end{eqnarray}
We also assume that the electronic density of states $\nu(\epsilon_F)$ is approximately a constant near the Fermi surface for $\delta q_{x}, q_{\perp} \ll k_F$. Leaving the details of calculation to appendix \ref{simdetails}, these simplifications lead to the following form of the spin-current for spin-up electrons flipping to spin-down ones. 
\begin{equation}
I_{spin,\uparrow} = \frac{\pi J^2 A_{\perp}  \nu(\epsilon_F)}{4}\int \frac{d\omega}{2\pi} \frac{d^{d-1}q_{\perp}}{(2\pi)^{d-1}}  \frac{ (V - \omega) }{1 - e^{-\beta(V - \omega)}} \; S_{-+}\left(\vec{q}_{\perp}, \omega \right)
\label{Ispinup}
\end{equation}
Analogous manipulations for the reverse process lead to 
\begin{equation}
I_{spin,\downarrow} = \frac{\pi J^2 A_{\perp}  \nu(\epsilon_F)}{4}\int \frac{d\omega}{2\pi} \frac{d^{d-1}q_{\perp}}{(2\pi)^{d-1}}  \frac{ (V + \omega) }{e^{\beta(V + \omega)}-1} \; S_{+-}\left(\vec{q}_{\perp}, \omega \right)
\label{Ispindown}
\end{equation}
These expressions make it transparent that as $T \rightarrow 0$, only up-spin electrons flipping to down-spin ones contribute the energy window $(0,V)$. The reverse process is always exponentially suppressed as there must be an energy gain of at least $V$ for a down-spin electron to flip to an up-spin one due to phase space constraints. The net spin-current is, as described in equation (\ref{spinc}), the difference of the above two currents.

This formalism can be extended to cases where the quasiparticle excitation energy has minima at large transverse momenta $\{\vec{Q}_{\perp}\}$ (with magnitude of $a^{-1}$ where $a$ is the microscopic lattice length-scale), provided the different $\vec{Q}_{\perp}$ are well-separated from each other. This is typically true for systems with quasiparticle bands, as the momenta difference between the band minima are of the order of $a^{-1}$. For example, cubic lattice antiferromagnets with a 2 dimensional boundary have spin-wave excitations about the ordering wave-vector $\vec{Q}^{AF}_{\perp} = \frac{\pi}{a}\left(0,1,1\right) $. Referring the reader to appendix \ref{simdetails} again for the details, here we just state the main result. The effect of inelastic scattering about large transverse momenta $\vec{Q}_{\perp}$ is to scale the spin-current by an overall $\mathcal{O}(1)$ angular factor $f_{ang}(k_F/Q_{\perp})$, so that equation (\ref{Ispinup}) for $I_{spin,\uparrow}$ is now modified to
\begin{equation}
I_{spin,\uparrow} =  \frac{\pi J^2 A_{\perp}  \nu(\epsilon_F)}{4} \sum_{\vec{Q}_{\perp}} f_{ang}(k_F/Q_{\perp})\int \frac{d\omega}{2\pi} \frac{d^{d-1}q_{\perp}}{(2\pi)^{d-1}}  \frac{ (V - \omega) }{1 - e^{-\beta(V - \omega)}} \; S_{-+}\left(\vec{q}_{\perp}, \omega \right)
\label{Ispinupfang}
\end{equation}
 where the angular factor, coming from kinematical constraints, is given by
\begin{equation}
f_{ang}(k_F/Q_{\perp}) = \int_{\stackrel{\hat{n}\cdot\hat{x} \geq 0 }{ k_F^2 (\hat{n}\cdot\hat{x})^2 + 2 k_F ( \vec{Q}_{\perp}\cdot\hat{n})\geq Q_{\perp}^2}} \frac{d\Omega}{S_{d-1}} \left( 1 + \frac{k_F (\hat{n}\cdot\hat{x})}{\left( k_F^2 (\hat{n}\cdot\hat{x})^2 + 2 k_F ( \vec{Q}_{\perp}\cdot\hat{n}) - Q_{\perp}^2 \right)^{1/2}} \right)
\label{fang}
\end{equation}
In equation (\ref{fang}), $S_{d-1}$ is the sphere in $\mathbb{R}^{d}$, and one can check that for $Q_{\perp} =0$ the angular factor reduces to unity, as desired. One can also check the limit $Q_{\perp} \gg k_F$, in which case scattering of the electron by $\vec{q}_{\perp} \approx \vec{Q}_{\perp}$ is excluded by phase space constraints and $f_{ang}(k_F/Q_{\perp}) \rightarrow 0$. Equation (\ref{Ispindown}) also undergo similar modifications, and putting these together we obtain our main result of this section 
\begin{eqnarray}
I_{spin} = \frac{\pi J^2 A_{\perp}  \nu(\epsilon_F)}{4} \sum_{\vec{Q}_{\perp}} f_{ang}(k_F/Q_{\perp})\int \frac{d\omega}{2\pi} \frac{d^{d-1}q_{\perp}}{(2\pi)^{d-1}} \left[ \frac{ (V - \omega) }{1 - e^{-\beta(V - \omega)}} \; S_{-+}\left(\vec{q}_{\perp}, \omega \right)  
 -  \frac{ (V + \omega) }{e^{\beta(V + \omega)}-1} \; S_{+-}\left(\vec{q}_{\perp}, \omega \right) \right] \notag \\
 \label{Ispin}
\end{eqnarray}

We once again carefully note that this formalism for extension of the spin-current calculation to a set of different $\{\vec{Q}_{\perp}\}$ works only when the different points are well-isolated in the Brillouin zone of spin-carrying excitations of the insulator. Physically, this implies that the different momentum patches (to which the electron is scattered) do not overlap with each other. If they start to overlap, then we would count the same final electron state multiple times and over-estimate the spin-current.  

\section{Spin current for ordered antiferromagnets}
\label{af}
In this section, we apply the formalism developed in section \ref{formalism} to calculate the spin-current from the metallic reservoir to an ordered collinear antiferromagnet, deep in the Neel phase. We assume $d = 3$, so that a symmetry-broken state can occur at $T > 0$. The results can also be generalized to $d=2$ at $T=0$. In the following subsections, we illustrate evaluation of the current with the simplest scenario - a cubic lattice antiferromagnet with ordering wave vector $\vec{Q}^{AF} = \frac{\pi}{a}\left(1, 1 ,1\right)$, so that $\vec{Q}^{AF}_{\perp} = \frac{\pi}{a}\left(0,1,1\right) $. We split our analysis into two subsections, corresponding to the Neel order pointing perpendicular and parallel to the spin-quantization axis in the metal, and add up the contributions due to elastic reflection from the static magnetic moments, and the inelastic reflection due to spin-wave excitations, to find the net spin-current. 

\subsection{Neel order perpendicular to spin quantization axis in the metal}
\subsubsection{Elastic contribution}
In order to contribute the elastic spin-flip scattering from the metal-antiferromagnet interface, we replace the fluctuating spin operators at the boundary by static moments, resembling the classical ground state. For Neel order along $\hat{y}$, which is normal to the spin-quantization axis $\hat{z}$ in the metal reservoir, we can write the Hamiltonian as
\begin{equation}
H_{int} = J \sum_{j} \vec{S}_e \cdot \vec{S}_j \, \delta ( \vec{x} - \vec{X}_j ) \rightarrow JS \sum_{j} S_y \; e^{-i\vec{Q}_{\perp}\cdot\vec{X}_j} \, \delta ( \vec{x} - \vec{X}_j )
\end{equation}
We use Fermi's golden rule again to find the rate of scattering of spin-flip scattering of electrons at the interface
\begin{equation}
R = 2\pi |\bra{\vec{k_2}, \downarrow} H_{int} \ket{\vec{k_1}, \uparrow}|^2 \delta(\epsilon_{\vec{k}_1} - \epsilon_{\vec{k}_2}) = \frac{\pi J^2}{4 L_x^2} \; \delta_{\vec{q}_{\perp},\vec{Q}_{\perp}} \;\delta\left(\epsilon_{\vec{k}_1} - \epsilon_{\vec{k}_1 - \vec{q}}\right)
\end{equation}
Following an analogous procedure of finding the spin current due to this scattering event, and summing over all initial and final states consistent with phase space restrictions, we arrive at the following expression for the elastic contribution $I_{spin}$ in terms of $f_{ang}\left(k_F/Q^{AF}_{\perp}\right)$
\begin{eqnarray}
  I^{el}_{spin,\uparrow} &=& f_{ang}\left(k_F/Q^{AF}_{\perp}\right)  \frac{\pi J^2 A_{\perp}}{8} \frac{\nu(\epsilon_F)V }{1- e^{-\beta V}} \\
  I^{el}_{spin,\downarrow} &=& f_{ang}\left(k_F/Q^{AF}_{\perp}\right)  \frac{\pi J^2 A_{\perp}}{8} \frac{\nu(\epsilon_F)V }{e^{\beta V}-1} \\
   I^{el}_{spin} &=&  I^{el}_{spin,\uparrow} -  I^{el}_{spin,\downarrow}  =  f_{ang}\left(k_F/Q^{AF}_{\perp}\right) \frac{\pi J^2 A_{\perp} \nu(\epsilon_F)}{8} V \label{IspinAFel}
\end{eqnarray}
Note that the elastic contribution to the current is proportional to the number of propagating modes at the Fermi surface, given by $\nu(\epsilon_F)V$. So this contribution is similar to what one would obtain by  using the Landauer formalism, as had been done for an analogous geometry by Takei {\em et al.\/}\cite{PhysRevB.90.094408}. 

\subsubsection{Inelastic contribution}
\label{afspinc}

The inelastic contribution can be directly evaluated by application of equation (\ref{Ispin}), as the ordered antiferromagnet deep in the Neel phase has spin-wave excitations that have minimum energy about $\vec{Q}_{\perp} = 0$ and $\vec{Q}_{\perp} =  \vec{Q}^{AF}_{\perp}$, which are well-separated in the insulator Brillouin zone. We work in the $T \rightarrow 0$ limit, which implies that the insulator is initially in its ground state. Therefore $\omega \geq 0$ in the dynamic structure factors, and we can drop the contribution from $I^{inel}_{spin,\downarrow}$ to the spin-current. 

We use the Holstein-Primakoff transformation to diagonalize the Hamiltonian and evaluate $S_{-+}\left(\vec{q}_{\perp}, \omega \right)$. Leaving the details to appendix \ref{afSF}, the dynamic structure factor in the small $|\vec{q}_{\perp}|$ and $T \rightarrow 0$ limit is given by (for $\omega >0$, setting $a=1$)
\begin{equation}
S_{-+}\left(\vec{q}_{\perp}, \omega \right) = \frac{\pi q_{\perp}}{8\sqrt{2}} \, \delta(\omega - v_s q_{\perp}) 
\label{sfafperp}
\end{equation}
where $v_s$ is the speed of spin-waves in the antiferromagnet. We can plug this back into equation (\ref{Ispin}), and we obtain the inelastic contribution to be 
\begin{equation}
I^{inel}_{spin} \stackrel{T \rightarrow 0}{=}  \frac{\pi J^2 A_{\perp}  \nu(\epsilon_F)}{4} \left[ 1+ f_{ang}\left(k_F/Q^{AF}_{\perp}\right) \right] \frac{V^4}{384\sqrt{2} \pi v_s^3} \label{IspinAFinel}
\end{equation}

We now add up the contributions from equations (\ref{IspinAFel}) and (\ref{IspinAFinel}) to find the net spin-current when the Neel order is perpendicular to the spin-quantization axis in the metal. 
\begin{equation}
I_{spin} \stackrel{T \rightarrow 0}{=}  \frac{\pi J^2 A_{\perp}  \nu(\epsilon_F)}{8} \left[ f_{ang}\left(k_F/Q^{AF}_{\perp}\right) V +  \left[ 1+ f_{ang}\left(k_F/Q^{AF}_{\perp}\right) \right] \frac{V^4}{192\sqrt{2} \pi v_s^3} \right]
\end{equation}

\subsection{Neel order parallel to spin quantization axis in the metal}
\subsubsection{Elastic contribution}
For Neel order along $\hat{z}$, which is normal to the spin-quantization axis $\hat{z}$ in the metal reservoir, we can write the Hamiltonian as
\begin{equation}
H_{int} = J \sum_{j} \vec{S}_e \cdot \vec{S}_j \, \delta ( \vec{x} - \vec{X}_j ) \rightarrow JS \sum_{j} S_z \; e^{-i\vec{Q}_{\perp}\cdot\vec{X}_j} \, \delta ( \vec{x} - \vec{X}_j )
\end{equation}
In this case, the Hamiltonian $H_{int}$ commutes with the $z$-component of the electron spin, and therefore cannot flip it. Therefore there is no elastic contribution to the spin-current. 

\subsubsection{Inelastic contribution}
For the inelastic contribution, we again use the $T \rightarrow 0$ limit of equation (\ref{Ispin}). The dynamic structure factor is evaluated in an analogous manner to the previous subsection \ref{afspinc}, and is essentially identical to equation (\ref{sfafperp}) barring a constant extra pre-factor. We find that the net spin current when the Neel vector is along the spin-quantization axis is given by
\begin{equation}
I_{spin} \stackrel{T \rightarrow 0}{=} I^{inel}_{spin} \stackrel{T \rightarrow 0}{=}  \frac{\pi J^2 A_{\perp}  \nu(\epsilon_F)}{4} \left[ 1+ f_{ang}\left(k_F/Q^{AF}_{\perp}\right) \right] \frac{V^4}{96\sqrt{2} \pi v_s^3} 
\end{equation}

\section{Spin current for systems with no magnetic order}
\label{non-mag}
In this section, we shall apply the formalism from section \ref{formalism} to evaluate the spin-current into states with no long range magnetic order. Some candidate phases for Mott insulators with unbroken spin-rotation symmetry are described by spin-half quasiparticles or spinons, coupled to an emergent gauge field. In the deconfined phase of the gauge field, the lattice symmetry is unbroken and the ground state is a spin liquid\cite{PhysRevB.45.12377}. The spinons can propagate as independent quasiparticles and carry a spin-current. In the confined phase, the ground state might spontaneously break translation symmetry of the lattice, resulting in a valence bond solid (VBS) state\cite{PhysRevB.42.4568} with short-range order. In this case, the low-lying excitations with non-zero spin are spin-triplets or triplons, which are gapped excitations that carry the spin current. 

\subsection{VBS states with triplon excitations}
At low energies, the structure factor will be dominated by single triplon excitations. Let us assume that the triplon has a gap $\Delta_T$ and a quadratic dispersion, so the dynamic structure factor can be approximated by 
\begin{equation}
S_{-+}\left(\vec{q}_{\perp}, \omega \right) \approx C \, \delta\left(\omega - \Delta_T - \gamma \vec{q}_{\perp}^{\;2} \right)
\end{equation}
Here we also assume that the prefactor $C$ is independent of $\omega$ and $\vec{q}_{\perp}$. Now we again use the $T \rightarrow 0$ limit of equation (\ref{Ispin}) to compute the spin-current. For a $d$ dimensional system with a $d-1$ dimensional boundary, we find that the spin-current is given by
\begin{equation}
 I_{spin} \stackrel{T \rightarrow 0}{=}  \frac{\pi J^2 A_{\perp} C S_{d-1} \gamma^{1 - d/2}  \nu(\epsilon_F)}{2(2\pi)^{d}d(d+1)}\, (V - \Delta_T)^{d/2+1} \, \Theta(V - \Delta_T)
\end{equation}
As expected, there is a threshold at $V = \Delta_T$, as energy conservation implies that no triplons can be excited when $V$ is less than the triplon gap. Above the cutoff, the spin-current has a power law behavior with voltage with an exponent that depends on the dimensionality $d$ of the system. For instance, in $d=3$, the exponent is $\frac{5}{2}$.

\subsection{Spin liquids with spinon excitations}
We first approach the problem analytically by using  a low energy effective theory to calculate the two-spinon structure factor. We use a mean-field approach where the spinons are free quasiparticles in the system, and have negligible coupling to other excitations which do not carry spin (like visons, which are vortices of the emergent gauge field). For a given spinon dispersion $\epsilon_{\vec{k}}$, the free-spinon Green's function in imaginary time is given by
\begin{equation}
G_s(\vec{k},i \omega_n) = \frac{1}{i\omega_n - \epsilon_{\vec{k}}}
\end{equation}
where $\omega_n$ is a Matsubara frequency which is determined by bosonic or fermionic statistics of the spinons. We can calculate the structure factor from the dynamic susceptibility 
$\chi_{-+}$, given by
\begin{eqnarray}
\chi_{-+}(\vec{q}_{\perp},i\omega_n) &=&  -\frac{1}{\beta \mathcal{V}} \sum_{\vec{k},i\Omega_n} 
G_s(-\vec{k},-i \Omega_n) G_s(\vec{k} + \vec{q}_{\perp},i \Omega_n + i \omega_n) \notag \\
& = & \int \frac{d^2k}{(2\pi)^2} \left( \frac{1 - n_B(\epsilon_{\vec{k}}) - n_B(\epsilon_{\vec{k} + \vec{q}_{\perp}})}{-i\omega_n + \epsilon_{\vec{k}} + \epsilon_{\vec{k} + \vec{q}_{\perp}}} \right)  \; \text{ (for bosonic spinons) }\notag \\
& \stackrel{T \rightarrow 0}{\rightarrow}  & \int \frac{d^2k}{(2\pi)^2} \frac{1}{(-i\omega_n + \epsilon_{\vec{k}} + \epsilon_{\vec{k} + \vec{q}_{\perp}})} 
\end{eqnarray} 
which, in turn, leads to the following result for the zero-temperature limit of the dynamic structure factor
\begin{eqnarray}
S_{-+}(\vec{q}_{\perp},\omega) &=& \frac{1}{1- e^{-\beta \omega}} \text{ Im}[\chi_{-+}(\vec{q}_{\perp},i\omega_n \rightarrow \omega + i \eta) ]  \notag \\
&\stackrel{T=0,\omega >0}{\rightarrow}& \lim_{T \rightarrow 0} \text{ Im}[\chi_{-+}(\vec{q}_{\perp},i\omega_n \rightarrow \omega + i \eta)] = \pi \int \frac{d^2k}{(2\pi)^2} \delta \left(\omega - \epsilon_{\vec{k}} + \epsilon_{\vec{k} + \vec{q}_{\perp}}\right)
\label{SF2spinon}
\end{eqnarray} 
Intuitively, this follows from the fact that spinons are always excited in pairs and they share the momentum transferred from the electron at the interface. At $T = 0$, the spin liquid is initially in its ground state, so we only have contributions from two spin-up spinons that have center of mass momentum $\vec{q}_{\perp}$. Equation (\ref{SF2spinon}) is the main result of this section, which we shall use to find the forms of the spin-current for certain spin-liquids with free-spinon bands in the mean-field picture, and then figure out how the spin-current scales with the spin-accumulation voltage $V$ for arbitrary spinon dispersions and dimensionality of the system. 

\subsubsection{Gapped spinons with quadratic bands}
Let us consider the case of gapped spin liquids in 2 dimensions with a spinon-gap $\Delta_s$, where the lowest spinon band has a quadratic dispersion about a minima at $\vec{k} = \vec{Q}_{\perp}$ with an effective mass of $m^{*}$, so that the spinon Green's function is given by
\begin{equation}
G_s(\vec{k},i\omega_n) = \frac{1}{i\omega_n - \Delta_s - \frac{(\vec{k}- \vec{Q}_{\perp})^2}{2m^*}}
\end{equation}
This is true for several ansatz spin liquid ground states\cite{PhysRevB.83.224413,KaiLi}, including, for instance, the $Q_1 = Q_2$ state of the $\mathbb{Z}_2$ spin liquid state on the Kagome lattice\cite{PhysRevB.45.12377}, where the gap and the effective mass are given in terms of the mean-field parameters $\lambda$ and $Q$, and the antiferromagnetic coupling between nearest neighbors $J_{AF}$ by
\begin{equation}
\Delta_s = \sqrt{\lambda^2 - 3 J_{AF}^2 Q^2}, \; \text{ and } \frac{1}{m^*} = \frac{3 J_{AF}^2 Q^2}{2\Delta_s}
\end{equation}
Equation (\ref{SF2spinon}) now leads to the following expression for the structure factor
\begin{eqnarray}
S_{-+}\left(\vec{q}_{\perp},\omega \right) &=&  \int \frac{d^2k}{(2\pi)^2} \, \delta \left(\omega - 2\Delta_s - \frac{(\vec{k}- \vec{Q}_{\perp})^2}{2m^*} - \frac{(\vec{k}+ \vec{q}_{\perp} - \vec{Q}_{\perp})^2}{2m^*} \right) \notag \\
& = & \frac{m^*}{4} \Theta\left(\omega - 2\Delta_s - \frac{\vec{q}_{\perp}^{\;2}}{4m^*}\right)
\label{2spinonSFQuadratic}
\end{eqnarray}  
In general, we may have several spinon bands with minima at different $\vec{Q}_{\perp}$ with the same gap $\Delta_s$, so we sum over all of them to find the net spin-current via equation (\ref{Ispin}) in the $T \rightarrow 0$ limit.
\begin{eqnarray}
I_{spin} &=& \frac{\pi J^2 A_{\perp}  \nu(\epsilon_F)}{4} \sum_{\vec{Q}_{\perp}} f_{ang}(k_F/Q_{\perp})\int \frac{d\omega}{2\pi} \frac{d^{d-1}q_{\perp}}{(2\pi)^{d-1}}  (V - \omega)  S_{-+}\left(\vec{q}_{\perp}, \omega \right) \notag \\
& = &  \frac{ \eta J^2 A_{\perp}  \nu(\epsilon_F) (m^{*})^2}{96\pi^2} \left( \sum_{\vec{Q}_{\perp}} f_{ang}(k_F/Q_{\perp}) \right) (V - 2\Delta_s)^3 \;  \Theta(V - 2\Delta_s) \notag \\
& = & \eta_2 \; (V - 2\Delta_s)^3 \;  \Theta(V - 2\Delta_s)
\end{eqnarray}
where we have absorbed all constant pre-factors in $\eta_2$ to explicitly show the dependence on $V$. As expected, there is a cutoff at twice the spinon gap, i.e, no spin current for $V \leq 2 \Delta_s$, and a power law behavior above the threshold. 

Note that in the calculation above, we assume that both spinons come from bands that have minima at identical $\vec{Q}_{\perp}$. However, even if they come from different bands, say one with minima at $\vec{Q}_{\perp,1}$ and the other with $\vec{Q}_{\perp,2}$, they will just contribute to add extra pre-factors of $f_{ang}\left(k_F/(|\vec{Q}_{\perp,1} + \vec{Q}_{\perp,2}|)\right)$ in the expression for the spin-current, but would not change either the threshold or the power law behavior. However, if the bands have different spinon-gaps, say $\Delta_{s,1}$ and $\Delta_{s,2}$ then we expect the spin-current to show a second threshold when the spin-accumulation voltage $V$ crosses $\Delta_{s,1} + \Delta_{s,2}$, as the scattering process then now excite spinons from both bands. 

\subsubsection{Gapless spinons at Dirac points}
Let us consider spin liquids described by gapless fermionic spinons at discrete Dirac points $\{\vec{Q}_{\perp}\}$ in the Brillouin zone. The spinon dispersion is then given in terms of the spinon velocity $v$ at a Dirac point at $\vec{Q}_{\perp}$ by 
\begin{equation}
G_s(\vec{k},i\omega_n) = \frac{1}{i\omega_n - v|\vec{k}- \vec{Q}_{\perp}|}
\end{equation}
This is again conjectured to be true for certain spin-liquid ansatz, for example, the $\pi$-flux state\cite{PhysRevB.37.3774} of the Heisenberg antiferromagnetic Hamiltonian on a $2d$ square lattice, which has been argued to be stable against $U(1)$ gauge fluctuations\cite{PhysRevB.70.214437}. We again use equation ($\ref{SF2spinon}$) to evaluate the structure factor.
\begin{eqnarray}
S_{-+}\left(\vec{q}_{\perp},\omega \right) &=&  \int \frac{d^2k}{(2\pi)^2} \delta \left(\omega - v|\vec{k}| - v|\vec{k} + \vec{q}_{\perp}| \right) \notag \\
& = & \frac{1}{8\pi v^2} \frac{\omega^2 - q_{\perp}^2/2}{\sqrt{\omega^2 - q_{\perp}^2}}\Theta\left(\omega - v |\vec{q}_{\perp}|\right)
\end{eqnarray}
We now use equation (\ref{Ispin}) to find the net spin-current for $T \rightarrow 0$. 
\begin{equation}
I_{spin} = \frac{J^2 A_{\perp}  \nu(\epsilon_F)}{960 \pi^2 v^2} V^5 = \eta_1 \, V^{5}
\end{equation}
where we have again absorbed all pre-factors in $\eta_1$ to make the $V$-dependence explicit. The current takes non-zero value for any $V > 0$, as there is no gap to a two-spinon excitation. We have evaluated the current for a single Dirac point, although extensions to multiple Dirac points with different velocities can be done in an exact analogy with the previous subsection, and will not affect the threshold or the exponent in the power law. 

\subsubsection{Generic spinon dispersions and spatial dimensions}
In this subsection, we are going to generalize the above results for given spinon dispersion in $d=2$ to generic dispersions and arbitrary space dimensions $d-1$ of the metal-insulator boundary using scaling arguments. Although this approach does not give us the exact-prefactors, it is sufficient to find out the characteristic dependence $I_{spin}$ on $V$.  We would require that the lowest spinon-band has minima at discrete points in the Brillouin zone, which are well-separated from each other.  We start off with gapless spin liquids with power law dispersions, and find that the exponent of $V$ is directly related to the power law in the dispersion and the dimensionality of the system. Our results easily generalize to gapped spin liquids.

Let the spinons have a dispersion given by 
\begin{equation}
 \epsilon(\vec{k}) = v_{\alpha} |\vec{k}|^{\alpha} 
\end{equation}
The two-spinon structure factor is proportional to an integral over the allowed phase space consistent with energy conservation.
\begin{equation}
 S_{-+}(\vec{q}_{\perp},\omega) \sim \int k^{d-2} dk \; d\Omega_{d-2}  \;\delta(\omega - v_{\alpha} |\vec{k}|^{\alpha} - v_{\alpha} |\vec{k} + \vec{q}_{\perp}|^{\alpha}) 
\end{equation}
The solutions for $k$ (when the delta function is non-zero) can be written in terms of a dimensionless scaling function $\Phi(v_\alpha q_{\perp}^{\alpha}/\omega) $ as 
\begin{equation}
k = q_{\perp}  \; \Phi(v_\alpha q_{\perp}^{\alpha}/\omega) 
\end{equation}
The delta function in $\omega$ can be rewritten as a delta function in $k$ as follows (in terms of another dimensionless function $\Phi_1$ which comes from the Jacobian)
\begin{equation}
\delta(\omega - v_{\alpha} |\vec{k}|^{\alpha} - v_{\alpha} |\vec{k} + \vec{q}_{\perp}|^{\alpha})  = \delta(k - q_{\perp}  \; \Phi(v_\alpha q_{\perp}^{\alpha}/\omega) ) / \left[v_{\alpha} q_{\perp}^{\alpha -1} \Phi_1(v_\alpha q_{\perp}^{\alpha}/\omega) \right] 
\end{equation}
Now we can see how the dynamic structure factor scales without explicitly evaluating the integral.
\begin{equation}
S_{-+}(\vec{q}_{\perp},\omega) \sim q_{\perp}^{d-1 - \alpha} \; \Psi(v_\alpha q_{\perp}^{\alpha}/\omega) 
\end{equation}
The dimensionless scaling function $\Psi$ must involve a theta function of the form $\Theta(\omega - \zeta v_\alpha q_{\perp}^{\alpha})$, where $\zeta$ is some arbitrary numerical constant that depends upon the exact dispersion. This follows from the fact that a large center of mass momentum will inevitably result in a large energy for the spinon pair which is precluded by energy conservation. Here, we are assuming that $\omega$ is small enough so that both the spinons come from the bottom of the band(s).

Finally, we turn to the $T \rightarrow 0$ limit of equation (\ref{Ispin}) again to find the spin current. 
\begin{equation}
I_{spin} \sim \int_{0}^{V} (V-\omega) d\omega \int  dq_{\perp} \; q_{\perp}^{d-2} \; d\Omega_{d-2}  S_{-+}(\vec{q}_{\perp},\omega)
\end{equation}
Because of the $\Theta$ function in $S_{-+}(\vec{q}_{\perp},\omega)$, the momentum integral is restricted to $q \leq (\omega/v_{\alpha})^{1/\alpha}$, so dimensional analysis tells us that 
\begin{equation}
\int  dq_{\perp} \; q_{\perp}^{d-2} \; d\Omega_{d-2} S_{-+}(\vec{q}_{\perp},\omega) \sim (\omega/v_{\alpha})^{(2d-2 - \alpha)/\alpha} 
\end{equation}
The integral over $\omega$ scales as $V^2$, so the final result after putting all this information together is
\begin{equation}
I_{spin} \sim V^2 \times V^{(2d-2 - \alpha)/\alpha}  = V^{1 + 2(d-1)/\alpha}
\end{equation}
As a check, let us see if the scaling matches the previous two exact calculations. In both cases, we have $d-1=2$. For the gapped $\mathbb{Z}_2$ spin liquid in the limit of the gap $\Delta_s \rightarrow 0$, we have $\alpha = 2$ and hence, $I_{spin} \sim V^{1 + 2(3-1)/2} = V^3$. For the gapless $U(1)$ spin liquid with $\alpha =1$, we have $I_{spin} \sim V^{1 + 2(3-1)/1} = V^5$.

For generalizing to gapped spin liquids with a spin gap of $\Delta_s$, all we need to do is make the following replacement in all the previous calculations:
\begin{equation}
 \omega \rightarrow \omega - 2 \Delta_s 
\end{equation}
This in turn tells us that the spin current is given by 
\begin{equation}
I_{spin} \sim (V - 2\Delta_s)^{1 + 2(d-1)/\alpha} \; \Theta(V - 2\Delta_s)
\label{scalingForm}
\end{equation}
Equation (\ref{scalingForm}) is the main result of this section. It shows that by measuring the spin current as a function of voltage, it is possible to deduce both the nature of the spin gap as well as the effective dispersion of the low energy excitations. Note that at the level of low-energy effective field theory, the current does not depend on the detailed structure of the lattice, but only on the effective continuum dispersion, as expected. 

\subsection{Numerical results for a model $\mathbb{Z}_2$ spin liquid state on the Kagome lattice}
In this section, we extend the previous results for a gapped $\mathbb{Z}_2$ spin liquid state via numerical calculations. As a model state, we choose the $Q_1 = Q_2$ ground state on the Kagome lattice, described by Sachdev\cite{PhysRevB.45.12377}. The reason for choosing this state for further investigation is that the dynamical structure factor measured in neutron scattering experiments on Herbertsmithite single crystals\cite{han2012fractionalized} is in good qualitative agreement with the calculations in the $Q_1 = Q_2$ ground state by Punk et. al\cite{punk2014topological}. 

Following Sachdev\cite{PhysRevB.45.12377}, we use a large $N$ expansion technique based on the symplectic group Sp(N).  To generalize of $S_{i}^{-}S_{j}^{+}$ to Sp(N), we just extract the part of the Sp(N) invariant scalar product $\vec{S}_i\cdot\vec{S}_j$\cite{PhysRevB.45.12377} that corresponds to $\frac{1}{2}S_{i}^{-}S_{j}^{+}$. In terms of the flavor indices $m$ of the Schwinger bosons that make up the spins, it can be written as
\begin{equation}
S_{i}^{-}S_{j}^{+} = \frac{1}{2N^2} \sum_{m_1,m_2}\left( b^{\dagger}_{im_1\downarrow}b_{im_2\uparrow}b^{\dagger}_{jm_2\uparrow}b_{jm_1\downarrow} + b^{\dagger}_{im_1\downarrow}b_{im_2\uparrow}b^{\dagger}_{jm_1\uparrow}b_{jm_2\downarrow}\right)
\end{equation}
Note that this reduces exactly to $S_{i}^{-}S_{j}^{+}$ of $SU(2)$ when we have a single flavor. To simplify the expression, we note that the $N$ flavors are decoupled in the $N= \infty$ mean-field theory, and each of the $N$ flavors has an identical Hamiltonian. Therefore, each flavor gives the same contribution, which just cancels off the extra factor of $N^2$, and we just need to calculate each term for a single flavor. The spinon operators that diagonalize the mean field Hamiltonian are linear in the $b$ and $b^{\dagger}$ operators, hence the correlation function factorizes as follows
\begin{equation}
\langle S_{i}^{-}S_{j}^{+} \rangle = \frac{1}{2} \left( \langle b^{\dagger}_{i\downarrow} b_{j\downarrow} \rangle \langle b_{i\uparrow} b^{\dagger}_{j\uparrow} \rangle+ \langle b^{\dagger}_{i\downarrow} b^{\dagger}_{j\uparrow} \rangle \langle b_{i\uparrow} b_{j\downarrow} \rangle \right)
\end{equation}
Moving to Fourier space and keeping only terms that give contributions to $\omega > 0$ after analytic continuation, we find that the dynamic susceptibility is given by
\begin{eqnarray}
\chi_{-+}(\vec{q}_{\perp},i\omega_n) = \frac{1}{2N_s}\sum_{\vec{k}, i\Omega_n} & \left[U_{jl}(-\vec{k})V_{jm}(\vec{k}+ \vec{q}_{\perp}) + V_{jl}(-\vec{k}) U_{jm}(\vec{k}+ \vec{q}_{\perp})\right] U_{il}^{*}(-\vec{k}) V_{im}^{*}(\vec{k}+ \vec{q}_{\perp}) \notag \\
& \times G_{l}(-\vec{k},-i \Omega_n) G_{m}(\vec{k}+ \vec{q}_{\perp},i \omega_n + i \Omega_n)
\label{KagomeChi}
\end{eqnarray} 
where $\vec{q}_{\perp}$ belongs to the extended Brillouin zone, $N_s$ is the total number of sites, $U, V$ are the Bogoliubov matrices that diagonalize the mean-field Hamiltonian, and we have implicitly summed over all sublattice indices $\{i,j,l,m\}$. We are going to use equation (\ref{KagomeChi}) to numerically evaluate the exact mean-field structure factor. As a side note, we mention that in the low energy limit, where $\vec{k}$ is close to the bottom of a spinon band $\vec{Q}_{\perp}$, and $\vec{q}_{\perp}$ is also small, so that the sum of the two spinon energies satisfies the energy constraint, we can approximate the elements of the $U$ and $V$ by their values at $\vec{Q}_{\perp}$, and then we recover the dynamic structure factor evaluated in equation (\ref{2spinonSFQuadratic}).

\begin{figure}[!h]
\centering
\includegraphics[scale=0.9]{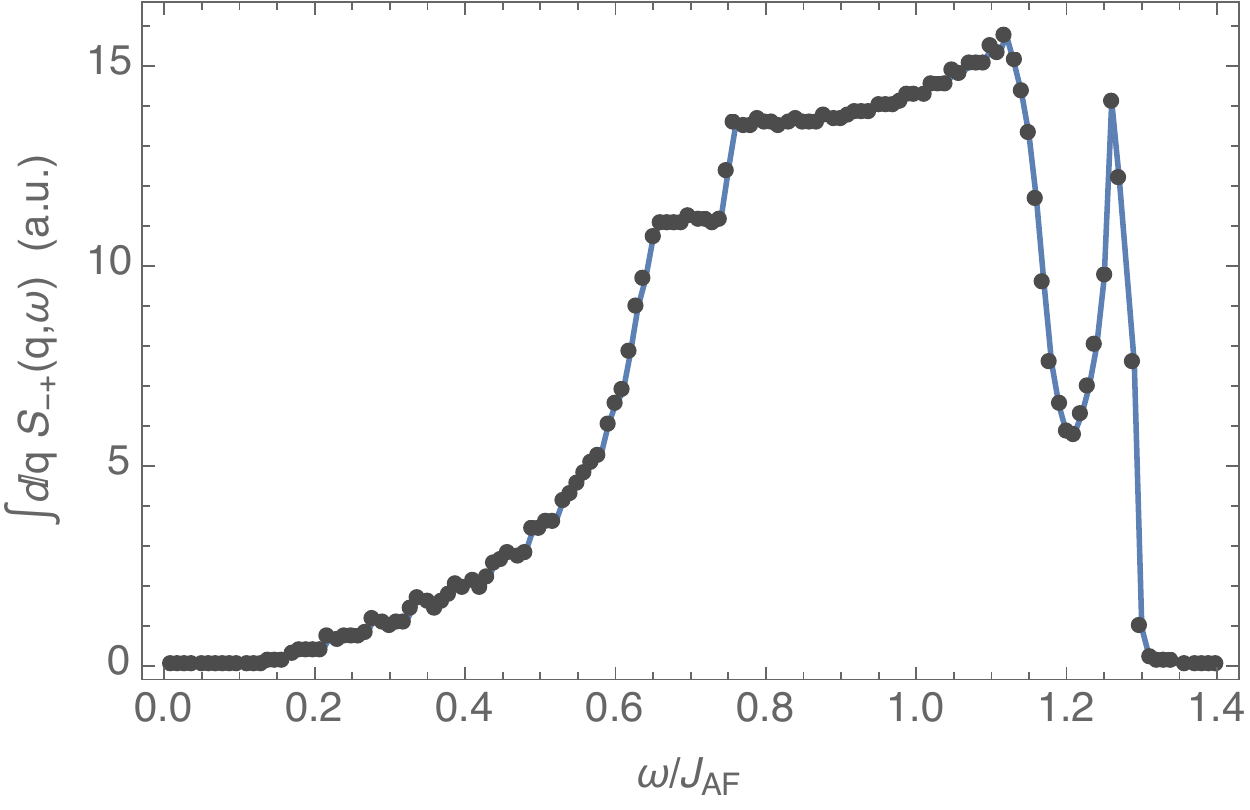}
\caption{Momentum integrated structure factor for the $Q_1 = Q_2$ ground state of the $ \mathbb{Z}_2$ spin liquid on the Kagome lattice}
\label{fig:QintegratedSF}
\end{figure}

We first plot the momentum-integrated structure factor $S_{-+}(\omega) = \frac{1}{N_s}\sum_{\vec{q}} S_{-+}(\vec{q},\omega)$ as a function of energy $\omega$ in Fig.~\ref{fig:QintegratedSF}. We assume mean-field parameters $\lambda = 0.695$ and $Q_1 = Q_2 = 0.4$ in the units of $J_{AF}$, which are \emph{not} self-consistently determined, and lead to a spinon gap of $\Delta_s \approx 0.5$.

We note two specific features, the jump at $\omega \approx 0.75$ and the peak at $\omega \approx 1.3$. Both these features can be understood using the band structure of the spinons for this ground state. The spinon spectra has a flat band with $\epsilon_{\vec{k}} = \lambda$, and once we have $\omega \geq \lambda + \Delta_s$, we can excite two spinons, one of them being at any momentum on the flat band. The second peak presumably comes from both spinons coming from the flat band, but is slightly smeared out by the Bogoliubov matrices and the finite width Lorenzian approximation for the delta function in the numerics. If we go up to energy scales of $V \approx J_{AF} \ll \epsilon_F$ (this is reasonable as $J_{AF} \approx 200 K$ for Herbertsmithite\cite{PhysRevLett.98.107204}, but typical $\epsilon_F \approx 10^{4}K$), we now can have contributions to the current at large values of $\delta q_x$ and $q_{\perp}$. In order to investigate the contributions properly, we need to numerically evaluate the spin-current starting with the $T \rightarrow 0$ limit equation (\ref{genIspinUp}). 

\begin{figure}[!h]
\centering
\includegraphics[scale=0.9]{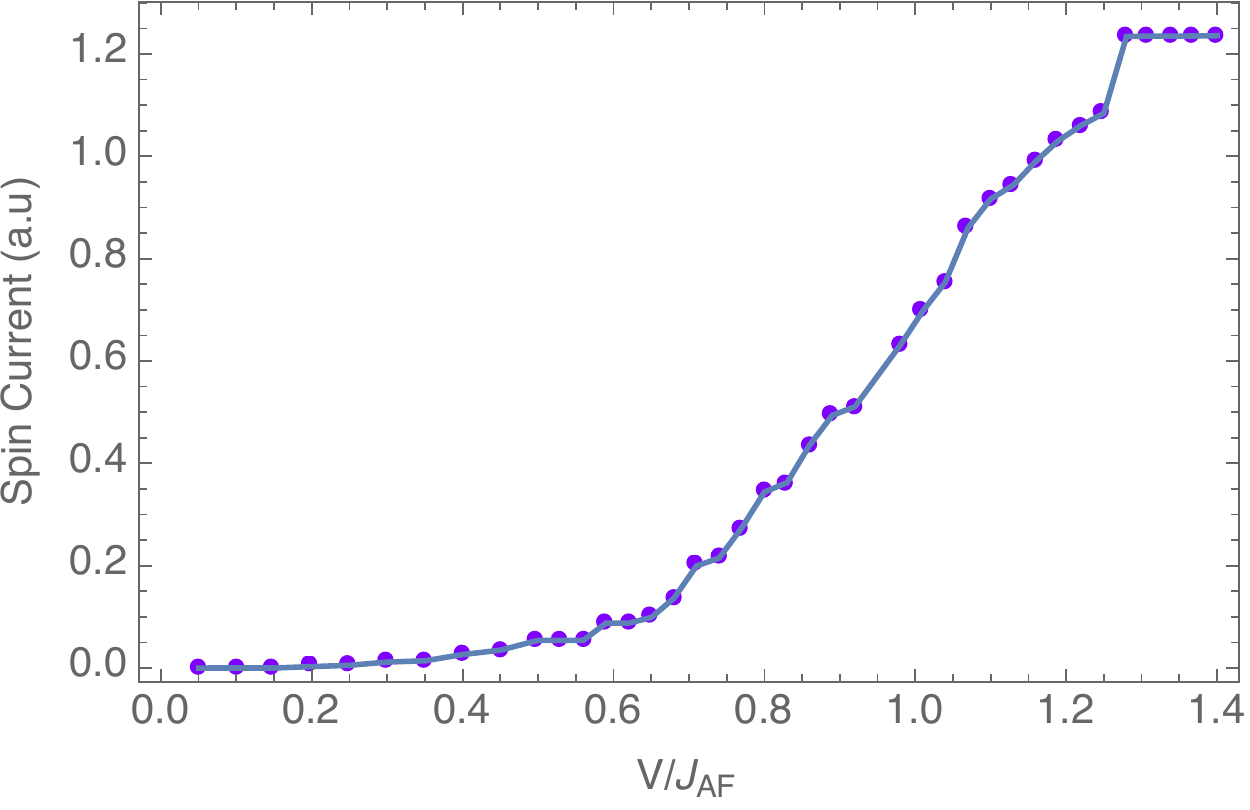}
\caption{Spin-current as a function of spin accumulation voltage for the $Q_1 = Q_2$ ground state of the $ \mathbb{Z}_2$ spin liquid on the Kagome lattice}
\label{fig:SpinCurrentQ1Q2}
\end{figure}

We next plot the spin-current, evaluated numerically, in Fig.~\ref{fig:SpinCurrentQ1Q2} as a function of the spin accumulation voltage $V$. The Fermi liquid parameters chosen for the plot below are $k_F = 2 $ (units of inverse lattice spacing), and $\epsilon_F/J_{AF} = 100$.

As expected, we observe the effects of the two features in the dynamic structure factor on the spin-current, which is roughly an integral over the structure factor. The step-like jump in the structure factor leads to a change in slope in the current around $V \approx 0.7$, and the spike leads to a step-like jump around $V \approx 1.3$, after which the current saturates. The observation of these two distinct features in the spin-current would be strong evidence in favor of the $Q_1 = Q_2$ $\mathbb{Z}_2$ spin liquid on the Kagome lattice. We note that the $Q_1 = -Q_2$ ground state\cite{PhysRevB.45.12377} does not have any flat spinon band, and is hence not expected to show any such feature in the spin-current.

\section{Conclusion and outlook}
In summary, we proposed the use of spin-currents as a gateway to probe the nature of excitations in magnetic insulators. Measurement of the spin-current as a function of the spin-accumulation voltage can throw light on the dispersion of the low-lying excitations and gap above the ground state. In particular, we showed at that in the zero temperature limit, the threshold and scaling of spin current with voltages may be used effectively to search for spin liquid ground states in magnetic insulators. Finally, we focused on a particular spin liquid ground state, which is a candidate state for Herbertsmithite\cite{punk2014topological}, and identified some broad features in the spin current which can help to identify that state. 

The spin-current is a valuable probe, because once injected into the insulator, the total spin is conserved in absence of spin-orbit coupling and random field impurities. We anticipate that it may be interesting to study how the presence of disorder in the interface, or the presence of non spin-carrying low-lying excitations in the insulator, which couple to the mobile spin-carrying modes, (for example, visons coupling to spinons\cite{punk2014topological} in spin liquids) affect the spin current.

\begin{acknowledgments}
Discussions with So Takei, Yaroslav Tserkovnyak and Amir Yacoby helped motivate this research.
We thank Debanjan Chowdhury, Soonwon Choi, and Bertrand Halperin for valuable discussions, and especially Matthias Punk for help with numerics. 
This research
was supported by the NSF under Grant DMR-1360789, the Templeton foundation, and MURI grant W911NF-14-1-0003 from ARO.
Research at Perimeter Institute is supported by the
Government of Canada through Industry Canada and by the Province of
Ontario through the Ministry of Research and Innovation.

\end{acknowledgments}

\appendix

\section{Details of spin-current calculations (from \ref{simplify})}
\label{simdetails}
We begin with the linearized energy transfer $\omega(\hat{n},\vec{q}_{\perp},\delta q_x)$ in equations [\ref{linearize}], and write the spin current from equation [\ref{genIspinUp}] as 
\begin{equation}
I_{spin,\uparrow} = \frac{\pi J^2 A_{\perp}}{4 m}\int_{k_{1x}>0} \frac{d^dk_1}{(2\pi)^d} \int_{q_x > k_{1x}} \frac{d^dq}{(2\pi)^d} n_{F}(\epsilon_{\vec{k}_1}) \left(1 - n_{F}(\epsilon_{\vec{k}_1} + V - \omega(\hat{n},\vec{q}_{\perp},\delta q_x)) \right) \: q_x \; S_{-+}\left(\vec{q}_{\perp}, \omega(\hat{n},\vec{q}_{\perp},\delta q_x) \right)
\label{simpeqn}
\end{equation}
The integral over $|\vec{k}_1|$ can now be evaluated, as everything else depends only on the direction $\hat{n}$ of the initial momentum, and the momentum transfer $\vec{q}$. Assuming that the density of states $\nu(\epsilon_F)$ is approximately a constant close to the Fermi surface, we have 
\begin{eqnarray}
&& \int_{k_{1x}>0} \frac{d^dk_1}{(2\pi)^d} n_{F}(\epsilon_{\vec{k}_1}) \left(1 - n_{F}(\epsilon_{\vec{k}_1} + V - \omega(\hat{n},\vec{q}_{\perp},\delta q_x)) \right) \notag \\
&\approx & \nu(\epsilon_F) \int_{\hat{n}\cdot\hat{x} >0} \frac{d\Omega}{S_{d-1}} \frac{V - \omega(\hat{n},\vec{q}_{\perp},\delta q_x)}{1 - e^{-\beta(V - \omega(\hat{n},\vec{q}_{\perp},\delta q_x))}}
\end{eqnarray}
We can further simplify equation [\ref{simpeqn}] by getting rid of $q_x$ in favor of $\omega$. For given $\vec{q}_{\perp}$ and $\hat{n}$, $d\omega = v_F (\hat{n}\cdot \hat{x}) d(\delta q_x)$ and $q_x \approx 2k_F (\hat{n}\cdot \hat{x})$, implying
\begin{equation}
\frac{dq_x \, q_x}{m} \approx -\frac{d(\delta q_x) 2k_F (\hat{n}\cdot \hat{x})}{m} = -2\, d\omega
\end{equation}
This is independent of the direction of initial momentum $\hat{n}$. Further, note that the constraint $q_x > k_F (\hat{n}\cdot \hat{x})$ is guaranteed to be satisfied by energy conservation, which requires small $\delta q_x$. By our assumption that $S_{-+}(\vec{q}_{\perp}, \omega)$ is insignificant for large $|\vec{q}_{\perp}|$, a change of energy due to large $\delta q_x$ cannot be offset by another due to large $|\vec{q}_{\perp}|$. The only problem arises when $\hat{n}\cdot\hat{x}$ is very small, but those are insignificant portions of the phase space that we can neglect. Therefore, all dependences of the $\vec{q}$-integral on $\hat{n}$ are removed, and this enables us to do the angular integral. Using $\int_{\hat{n}\cdot\hat{x} >0} \frac{d\Omega}{S_{d-1}} = \frac{1}{2}$, we recover the simplified expression stated in equation [\ref{Ispinup}] 
\begin{equation}
I_{spin,\uparrow} = \frac{\pi J^2 A_{\perp}  \nu(\epsilon_F)}{4}\int \frac{d\omega}{2\pi} \frac{d^{d-1}q_{\perp}}{(2\pi)^{d-1}}  \frac{ (V - \omega) }{1 - e^{-\beta(V - \omega)}} \; S_{-+}\left(\vec{q}_{\perp}, \omega \right)
\end{equation}
The calculation for $I_{spin,\downarrow}$ (equation [\ref{Ispindown}]) is analogous, with the only change coming from the different occupancies of the initial and final states.

We now discuss the case when the dynamic structure factor has a minima at large transverse momentum $\vec{Q}_{\perp}$. The trick is to note that although the momentum transfer can be large, the energy transfer at low temperatures is always small, i.e, $\omega \lesssim V \ll \epsilon_F$. Therefore, we can expand in small parameters about the point of elastic scattering. To do so, we first solve for a longitudinal momentum transfer $q_{x0}$ which satisfies $\epsilon_{\vec{k}_1} = \epsilon_{\vec{k}_1 - \vec{Q}}$, where $\vec{Q} = q_{x0} \hat{x} + \vec{Q}_{\perp}$.
\begin{eqnarray}
\epsilon_{\vec{k}_1} - \epsilon_{\vec{k}_1 - \vec{Q}} & = & 2k_F\left(\vec{Q}_{\perp}\cdot\hat{n} + q_{x0} (\hat{n}\cdot\hat{x}) \right) - (Q_{\perp}^2 + q_{x0}^2)= 0 \notag \\
\implies q_{x0} & = & k_F \, (\hat{n}\cdot\hat{x}) + \left( k_F^2 (\hat{n}\cdot\hat{x})^2 + 2 k_F ( \vec{Q}_{\perp}\cdot\hat{n}) - Q_{\perp}^2 \right)^{1/2}
\end{eqnarray}
The contraint $q_{x0} \geq k_F \, (\hat{n}\cdot\hat{x})$, required for reflection, implies that only the positive square root can contribute. As $q_x$ is real, only some values of $\hat{n}$ are relevant. Specifically, we require
\begin{equation}
 k_F^2 (\hat{n}\cdot\hat{x})^2 + 2 k_F ( \vec{Q}_{\perp}\cdot\hat{n})  \geq Q_{\perp}^2 
\end{equation}
We need to evaluate the angular integral over angular regions of the Fermi surface consistent with the above constraint. If $Q_{\perp}$, the magnitude of the transverse scattering wave vector, is too large compared to the Fermi momentum $k_F$, then there is no scattering consistent with energy conservation, and hence there is no spin current due to this process.  
 
Now, we can expand about the solution for elastic scattering for small $\omega$, and keep only linear terms in $\delta q_x$ and $\vec{q}_{\perp}$.
\begin{eqnarray}
\vec{q} & = & (q_{x0} - \delta q_x) \hat{x} + \vec{Q}_{\perp} - \vec{q}_{\perp} \notag \\
\omega &  = & \frac{1}{m}\left[ \left(q_{x0}- k_F (\hat{n}\cdot\hat{x})  \right)\delta q_x  + \left(\vec{Q}_{\perp} - k_F \hat{n}\right)\cdot\vec{q}_{\perp} \right] +  \mathcal{O}(\delta q_x^2, q_{\perp}^2) \notag \\
&& \text{ where we used } 2k_F\left(\vec{Q}_{\perp}\cdot\hat{n} + q_{x0} (\hat{n}\cdot\hat{x}) \right) - (Q_{\perp}^2 + q_{x0}^2)= 0 
\end{eqnarray}

We revert to our previous formalism, and replace the integral over $q_x$ by an integral over $\omega$, with the only change being in the pre factor appearing the angular integral. For fixed $\vec{q}_{\perp}$  and $\hat{n}$, $d\omega =  \frac{1}{m} \left(q_{x0} - k_F (\hat{n}\cdot\hat{x}) \right)\delta q_x $, and $q_x \approx q_{x0}$, implying that
\[ \frac{dq_x\, q_x}{m_e} \approx -\frac{d(\delta q_x) q_{x0}}{m_e} = d\omega \left( \frac{q_{x0}}{k_F (\hat{n}\cdot\hat{x}) - q_{x0}} \right) = - \left( 1+ \frac{k_F (\hat{n}\cdot\hat{x})}{\left( k_F^2 (\hat{n}\cdot\hat{x})^2 + 2 k_F ( \vec{Q}_{\perp}\cdot\hat{n}) - Q_{\perp}^2 \right)^{1/2}}\right) d\omega\]
Note that for $|\vec{Q}_{\perp}| \ll k_F$, we get back our previous result which corresponds to scattering at $\vec{Q}_{\perp}  = 0$. This acts as a check on the above calculation, and also shows that the calculation can be generalized as long as we have low energy excitations in the spin-system about a set of isolated points in momentum space which are well-separated in the Brillouin zone. 

The factor multiplying $d\omega$ will change the result of the angular integral over initial momenta, but the remaining calculation remains unchanged, and we have
\begin{eqnarray}
 I^{\vec{Q}_{\perp}}_{spin,\uparrow} &=& \frac{\pi J^2 A_{\perp}  \nu(\epsilon_F)}{4}  \int_{ k_F^2 (\hat{n}\cdot\hat{x})^2 + 2 k_F ( \vec{Q}_{\perp}\cdot\hat{n})\geq Q_{\perp}^2 } \frac{d\Omega}{S_{d-1}} \left( 1 + \frac{k_F (\hat{n}\cdot\hat{x})}{\left( k_F^2 (\hat{n}\cdot\hat{x})^2 + 2 k_F ( \vec{Q}_{\perp}\cdot\hat{n}) - Q_{\perp}^2 \right)^{1/2}} \right) \notag \\
&& \times \int \frac{d\omega}{2\pi} \frac{d^{d-1}q_{\perp}}{(2\pi)^{d-1}}  \frac{ (V - \omega) }{1 - e^{-\beta(V - \omega)}} \; S_{-+}\left(\vec{q}_{\perp}, \omega \right) \notag \\
& = & \frac{\pi J^2 A_{\perp}  \nu(\epsilon_F)}{4} f_{ang}(k_F/Q_{\perp}) \int \frac{d\omega}{2\pi} \frac{d^{d-1}q_{\perp}}{(2\pi)^{d-1}}  \frac{ (V - \omega) }{1 - e^{-\beta(V - \omega)}} \; S_{-+}\left(\vec{q}_{\perp}, \omega \right) 
\end{eqnarray}
where $f_{ang}(k_F/Q_{\perp})$ is the angular integral referred to in equation [\ref{fang}]. Typically, $k_F$ and $Q_{\perp}$ have the same order of magnitude, and then the angular integral is an overall factor of $\mathcal{O}(1)$ (the exact value is determined by the constraints set by the ordering wave vector $\vec{Q}_{\perp}$).

Taking into account that there can be multiple such minima in the dynamic structure factor at large finite momenta $\{ \vec{Q}_{\perp}\}$, and scattering to momenta patches around these minima are independent as long as the minima are well-separated, we arrive at equation [\ref{Ispinupfang}], stated below for the sake of completeness.
\begin{equation}
I_{spin,\uparrow} =  \frac{\pi J^2 A_{\perp}  \nu(\epsilon_F)}{4} \sum_{\vec{Q}_{\perp}} f_{ang}(k_F/Q_{\perp})\int \frac{d\omega}{2\pi} \frac{d^{d-1}q_{\perp}}{(2\pi)^{d-1}}  \frac{ (V - \omega) }{1 - e^{-\beta(V - \omega)}} \; S_{-+}\left(\vec{q}_{\perp}, \omega \right)
\end{equation}
The expression for $I_{spin,\downarrow}$ follows in exact analogy to the above calculation.

\section{$S_{-+}\left(\vec{q}_{\perp}, \omega \right)$ for an antiferromagnetic interface (from \ref{afspinc})}
\label{afSF}
We evaluate the dynamic structure factor for a Neel-ordered state on a $d$-dimensional cubic lattice using the Holstein-Primakoff transformation. First, let us consider the case when the Neel vector points parallel to the spin-quantization axis in the metal (chosen to be $\hat{z}$). We have up-spins on sub lattice A and down-spins on sub lattice B, with total number of spins being be $2N$, and the coordination number of each spin is $z = 2d$. Therefore we define
\begin{eqnarray}
i \in A, \; S_i^{-} = a_i^{\dagger} (2S - a_i^{\dagger}a_i)^{1/2}; \; 
S_i^{+} = (2S - a_i^{\dagger}a_i)^{1/2} \, a_i , \text{ and } S_{i}^{z} = S - a_i^{\dagger}a_i \notag \\
i \in B, \; S_i^{+} =  b_i^{\dagger} (2S - b_i^{\dagger}b_i)^{1/2}; \; S_i^{+} = (2S - b_i^{\dagger}b_i)^{1/2} \, b_i , \text{ and } S_{i}^{z} = -S + b_i^{\dagger}b_i
\end{eqnarray}
and do an expansion in $1/S$. The Heisenberg Hamiltonian $H_{AF} = J_{AF} \sum_{<ij>} \vec{S}_i \cdot \vec{S}_j$ can be written in terms of the Holstein Primakoff bosons as 
\begin{equation}
 H_{AF} = - J_{AF}NS^2 z + J_{AF} Sz \sum_{\vec{k}}[a^{\dagger}_{\vec{k}} a_{\vec{k}} + b^{\dagger}_{\vec{k}} b_{\vec{k}} + \gamma_{\vec{k}}(a_{\vec{k}} b_{-\vec{k}} + a^{\dagger}_{\vec{k}}b^{\dagger}_{-\vec{k}} )] + \mathcal{O}(S^{0}), 
\; \text{ where } \gamma_{\vec{k}} = \frac{1}{z}\sum_{\delta \in n.n} e^{i\vec{k}\cdot\vec{\delta}}
\end{equation}
This can be diagonalized by a Bogoliubov transformation, using 
\begin{eqnarray}
a_{\vec{k}}  = u_{\vec{k}} \alpha_{\vec{k}} + v_{\vec{k}} \beta^{\dagger}_{-\vec{k}} \; ,&\; \; 
b_{\vec{k}}  = u_{\vec{k}} \beta_{\vec{k}} + v_{\vec{k}} \alpha^{\dagger}_{-\vec{k}} \notag \\
\text{with } u_{\vec{k}} = u_{-\vec{k}} = \text{cosh}(\theta_{\vec{k}}), & \; v_{\vec{k}} = v_{-\vec{k}} = \text{sinh}(\theta_{\vec{k}}) \; \text{ and tanh}(2\theta_{\vec{k}}) = - \gamma_{\vec{k}}
\end{eqnarray}
The Hamiltonian is diagonal in terms of the Bogoliubov quasiparticles, 
\[ H_{AF} = - J_{AF}NS^2z - J_{AF}NSz + \sum_{\vec{k}} E_{\vec{k}}(\alpha^{\dagger}_{\vec{k}}\alpha_{\vec{k}} + \beta^{\dagger}_{\vec{k}}\beta_{\vec{k}} + 1), \; E_{\vec{k}} = J_{AF}Sz\sqrt{1 -  \gamma^2_{\vec{k}}} \]
$S_{-+}(\vec{q}_{\perp}, \omega)$ may now be calculated from definition using the expression for the spin operators in terms of in terms of the quasiparticle creation and annihilation operators. After some algebra, we find 
\begin{equation}
S_{-+}(\vec{q}_{\perp},\omega) = 2\pi S(u_{\vec{q}_{\perp}} +  v_{\vec{q}_{\perp}})^2\left[ \delta(\omega - E_{\vec{q}_{\perp}})(1 + n(\beta_{\vec{q}_{\perp}})) + \delta(\omega + E_{\vec{q}_{\perp}})n(\alpha_{-\vec{q}_{\perp}}) \right]
\end{equation}
At $T=0$, only the delta function with positive $\omega$ contributes to the spin current as there are on quasiparticles initially in the system. For low momenta, we have
\begin{equation}
E_{\vec{q}_{\perp}} \approx v_s |\vec{q}_{\perp}|, \text{ and }
(u_{\vec{q}_{\perp}} +  v_{\vec{q}_{\perp}})^2 = \text{cosh}(2\theta_{\vec{q}_{\perp}}) + \text{sinh}(2\theta_{\vec{q}_{\perp}}) = \sqrt{\frac{1- \gamma_{\vec{q}_{\perp}}}{1+ \gamma_{\vec{q}_{\perp}}}} = \frac{q_{\perp}}{2\sqrt{d}}
\label{lowPlimitAF}
\end{equation}
which leads to the following expression for the dynamic structure factor for the $T=0$ antiferromagnet 
\begin{equation}
S_{-+}(\vec{q}_{\perp},\omega) = \frac{\pi S q_{\perp}}{\sqrt{d}} \; \delta(\omega - v_s q_{\perp})
\end{equation}
For the case when the Neel order is perpendicular to the spin quantization axis in the metal, we assume that spins on sub lattice $A$ are pointing in the $\hat{y}$ direction and the spins on sub lattice $B$ are pointing in the $-\hat{y}$ direction. In this case, we can still use the Holstein-Primakoff representation of spins after doing a $\pi/2$ rotation of our coordinate system with respect to the x axis. In the rotated coordinate system $XYZ$ we have $X = x, Y = -z,$ and $Z = y$. Remembering that our original definitions of $S^{\pm}$ were with respect to the old axes, let us denote our spin operators by $\Sigma$ in the new set of axes. Then
\begin{equation}
S^{\pm} = S^{x} \pm i S^{y} = \Sigma^{X} \pm i \Sigma^{Z} 
\end{equation}
We can now express these in terms of the usual Holstein Primakoff bosons, and after some algebra, find the following dynamic structure factor in the large-S approximation
\begin{equation}
 S_{-+}(\vec{q}_{\perp},\omega) =  \frac{2\pi S}{4} (u_{\vec{q}_{\perp}} +  v_{\vec{q}_{\perp}})^2\left[ \delta(\omega - E_{\vec{q}_{\perp}})(2 + n(\alpha_{\vec{q}_{\perp}}) + n(\beta_{\vec{q}_{\perp}})) + \delta(\omega + E_{\vec{q}_{\perp}})(n(\alpha_{-\vec{q}_{\perp}}) + n(\beta_{-\vec{q}_{\perp}})) \right]
\end{equation}
Using the low momentum limit from [\ref{lowPlimitAF}] and taking $T \rightarrow 0$, we arrive at the expression in equation [\ref{sfafperp}].

\bibliography{paper1_bib}

\end{document}